# *Gravitational oscillations of a liquid column*

*Élise Lorenceau*[1], *David Quéré*[1], *Jean-Yves Ollitrault*[2] *& Christophe Clanet*[3]

*(1) Laboratoire de Physique de la Matière Condensée, URA 792 du CNRS, Collège de France, 75231 Paris Cedex 05, France*
*(2) Service de Physique Théorique, CE Saclay, 91191 Gif-sur-Yvette Cedex, France*
*(3) Institut de Recherche sur les Phénomènes Hors Equilibre, UMR 6594 du CNRS, 49 rue F. Joliot-Curie, BP 146, 13384 Marseille, France.*

We report gravity oscillations of a liquid column partially immersed in a bath of liquid. We stress in particular some peculiarities of this system, namely (i) the fact that the mass of this oscillator constantly changes with time; (ii) the singular character of the beginning of the rise, for which the mass of the oscillator is zero; (iii) the sources of dissipation in this system, which is found to be dominated at low viscosity by the entrance (or exit) effects, leading to a long-range damping of the oscillations. We conclude with some qualitative description of a second-order phenomenon, namely the eruption of a jet at the beginning of the rise.

## 1. Experiment

A cylindrical glass pipe, closed at its top, is partially immersed in a large bath of liquid. The experiment consists of opening the pipe, and recording the height $Z$ of the liquid column as a function of time $T$ (Figure 1). The pipe has a centimetric radius $R$ (which allows us to neglect capillary effects), and a total length of about one meter. We note $H$ the depth of immersion, and $h$ the level of liquid inside the tube before opening. This parameter can be adjusted thanks to a syringa, with which we can add either some liquid or some air before opening the top. We are interested here in liquids of low viscosity $\eta$ (such as water or hexane), so that the motion of the liquid is dominated by inertia and gravity, leading to numerous oscillations of the liquid column.

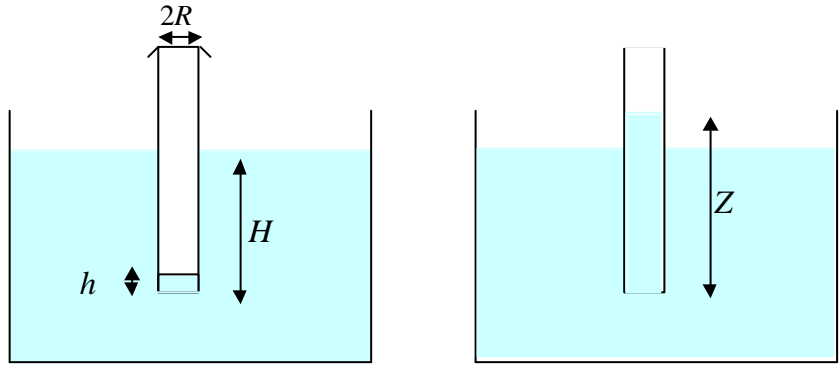

*Figure 1: Sketch of the experiment: a) before opening the top (T < 0); b) when the motion takes place (T > 0).*

Figure 2 shows typical observations of the column height as a function of time, obtained thanks to a high speed camera (~ 1200 frames per second). For this particular experiment, the initial depth was $H = 30$ *cm*, the tube radius $R = 1$ *cm*, and the initial height of liquid inside the column $h = 3$ *mm*. The liquid was hexane, of density $\rho = 660$ *kg/m³* and viscosity $\eta = 0.39$ *mPa.s*. Both the height and the time are made dimensionless in Fig. 2, where they are respectively normalized by the natural length and time scales, namely $H$ and $\sqrt{H/g}$, noting $g$ the acceleration of gravity. We note $z$ and $t$ these reduced quantities.

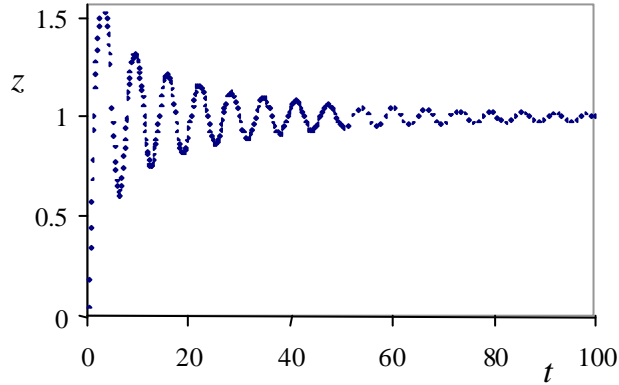

*Figure 2: Height of the liquid column versus time, for a glass tube (R = 1 cm) partially immersed at a depth H = 30 cm in hexane. Initially (t = 0), the tube is empty. The height is normalized by H, and the time by $(H/g)^{1/2}$.*

We can observe in Fig. 2 several features, on which we shall base our discussions: (i) the height first quickly increases (the typical velocity at the beginning is 170 cm/s), and reaches a maximum $z_M = 1.52 \pm 0.01$ for $t = 3.0 \pm 0.5$; (ii) then, many oscillations are observed, before approaching the final equilibrium height $z = 1$; their damping is not exponential, since the ratio between two successive maxima of the function $(z - 1)$ is not a constant (the four first ratios are respectively 0.61, 0.68, 0.73 and 0.77, and increase with time); (iii) a pseudo-period can be deduced from the data, which is $6.3 \pm 0.5$; this period is quite well defined for the first oscillations, but slightly increases (of typically less than 5 %) at longer times.

We shall first describe the principal characteristics of a model recently proposed to analyze the non-linear oscillations of a liquid column. Then, we shall discuss different effects such as speed of invasion, initial acceleration of the fluid, and damping. We shall conclude with some qualitative observations related to local properties of the flow.

## 2. Model

A model was recently proposed to describe the motion due to capillary forces of a wetting liquid inside a small vertical tube initially empty ($H = 0$), in the inertial regime.[1] Then, the force acting on the liquid column writes: $F = 2\pi R\gamma - \rho g\pi R^2 Z$ (noting $\gamma$ the liquid surface tension). Here, the tube radius is much larger than the (millimetric) capillary length, so that capillary forces can be neglected and replaced by the hydrostatic pressure as a driving force. Hence, the force F acting on the liquid column is found to exhibit a structure very similar to the one in the capillary problem:

$$F = \rho g\pi R^2 (H-Z) \qquad (1)$$

It is very instructive to consider first a situation without any source of dissipation. Then, the total energy E of the column is the sum of the kinetic energy and the potential energy U which can be integrated from F (taking $U = 0$ for $Z = 0$). Hence it writes:

$$E = \tfrac{1}{2}\rho\pi R^2 Z\dot{Z}^2 + \tfrac{1}{2}\rho g\pi R^2 Z^2 - \rho g H\pi R^2 Z \qquad (2)$$

In dimensionless variables (and scaling the mass by $\rho\pi R^2 H$), it reads:

$$e = \tfrac{1}{2}z\dot{z}^2 + \tfrac{1}{2}z^2 - z \qquad (3)$$

Considering *e* as a constant with time, eq. (3) can be integrated, which leads to parabolic oscillations of equation: $z(t) = \sqrt{2}\, t(1 - t/4\sqrt{2})$, supposing $z(0) = 0$. The maximum, reached at $t = 2\sqrt{2}$, is $z = 2$, far above the maximum observed in the experiment in Fig. 2. Furthermore, this parabolic behavior is periodic (with this peculiarity, which is that when *z* comes back to zero, the velocity is maximum, but there is no inertia because there is no mass; thus, the liquid column can bounce) – once again, in contradiction with the observed damping of the oscillations.

Thus, the second step consists of analyzing the possible causes of dissipation in the system. We could try to incorporate the viscous dissipation along the wall of the tube, but this should be negligible at short time – *i.e.* at a time scale smaller than $\rho R^2/\eta$, the characteristic time for setting a Poiseuille profile in the tube. This time in these experiments is very large, typically $10^2$ to $10^3$ in our dimensionless units. The negligible influence of viscosity at short time was confirmed by doing the same experiment with water (three times as viscous as hexane), for which we found exactly the same positions for the five first maxima and five first minima (within 1 % in error).

In classical textbooks[2], one can find that a second cause of dissipation for a liquid of very small viscosity is the singular pressure loss at the tube entrance (if the liquid rises) or exit (if the liquid goes down). This pressure loss is due to the difference of radii between the tank (of huge radius) and the tube (of much smaller radius): because of the abrupt contraction between both, some eddies appears at the entrance of the tube, dissipating a certain amount of energy. This pressure loss is classically evaluated by applying the Bernoulli equation (based on the conservation of energy) and the Euler equation (based on the momentum equation) to the liquid column, which does not lead to the same result[2]. The difference between these results is precisely the pressure loss. When the ratio of the tube radius to the tank radius is close to zero (in our experiment, this ratio is of order of $10^{-2}$), this pressure loss $\Delta P$ has a very simple expression[2]:

$$\Delta P = \tfrac{1}{2} \rho \dot{Z}^2 \qquad (4)$$

This pressure loss is positive and is simply equal to the kinetics energy of a slice d$Z$ of the column. The associated energy loss is negative, and has a different sign depending on whether the liquid is going up (d$Z > 0$) or down (d$Z < 0$). In dimensionless quantities, the energy loss thus writes:

$$de = -\frac{1}{2} \dot{z}^2 dz \qquad (5a)$$

when the liquid rises (d$z > 0$), and:

$$de = \frac{1}{2} \dot{z}^2 dz \tag{5b}$$

when it falls ($dz < 0$).

Differentiating eq. 3 (and taking no account for the viscous dissipation along the wall), we find two different equations, depending on the direction of the motion:

$$z\ddot{z} + \dot{z}^2 = 1 - z \quad \text{for } dz > 0 \tag{6a}$$
$$z\ddot{z} = 1 - z \quad \text{for } dz < 0 \tag{6b}$$

These equations can be simply derived by expressing Newton's law of dynamics, taking into account the fact that the system is open. For a system of mass M and velocity V driven by a force F, it can be written:

$$\frac{d}{dt} MV = F + \dot{M} V_o \tag{7}$$

where $V_o$ is the velocity with which the mass joins (or leaves) the system. For the rise, $\dot{M}V_o$ is close to zero (velocity field related to a punctual sink), and eq. 7 turns out to be identical to eq. 6a (after making dimensionless the different variables). During the descent, a liquid jet is expelled into the reservoir and $V_o$ is V, which leads to eq. 6b.

The energy loss associated with eq. 7 can be calculated in a very general way. The energy E is $1/2 MV^2 + U$, noting U the potential from which the force F can be derived ($F = - dU/dZ$). Using eq. 7, the way the energy varies as a function of time can be deduced. We find:

$$\frac{dE}{dt} = -\frac{1}{2} \dot{M} V^2 \left[ 1 - \frac{2V_o}{V} \right] \tag{8}$$

which leads to a unique expression for both the rise ($\dot{M} > 0$ and $V_o = 0$) and the descent ($\dot{M} < 0$ and $V_o = V$):

$$\frac{dE}{dt} = -\frac{1}{2}|\dot{M}|V^2 \tag{9}$$

Eq. 9 is found to be identical to eq. 5a and 5b. It expresses more generally the energy loss related with an entrained mass (dE/d$t$ = 0 if $\dot{M}$ = 0). It thus concerns similar questions such as the bursting of a soap film[3] or even the academic problem of a rope wound on a pulley and drawn by a constant weight.

Equations 6a and 6b can eventually be integrated once, introducing two constants $A$ and $B$:

$$\frac{1}{2}z^2\dot{z}^2 + \frac{1}{3}z^3 - \frac{1}{2}z^2 = A \tag{10a}$$

$$\frac{1}{2}\dot{z}^2 + z - \ln z = B \tag{10b}$$

If $z = 0$ at $t = 0$, the constant $A$ is zero, and equation 10a can be integrated once again, which provides the trajectory of the liquid column:

$$z(t) = t\left(1 - \frac{t}{6}\right) \tag{11}$$

Thus, the beginning of the rise should be linear ($z \sim t$, for $t \ll 6$), before the weight makes the velocity smaller and the trajectory parabolic. The maximum is reached for $t = 3$, and is found to be $z_M = 1.5$. This latter point is in close agreement with the data displayed in Fig. 2, which stresses that indeed energy loss is present in the system, even at short time.

We now present more detailed experiments, in order to discuss more carefully the details of the model.

## 3. Discussion

### 3.1. Constant velocity regime

Initially the beginning of the rise is linear, which can be explained by balancing inertia with the pressure force ($\rho g H \pi R^2$) exerted on the liquid column. This behavior is reminiscent of similar systems with a mass varying linearly with $z$, and driven by a constant force and resisting inertially. This indeed leads to a constant velocity, as observed for the retraction of a liquid sheet[4], the bursting of a soap film[3,5], the dewetting of a film of small viscosity[6] and the first steps of capillary rise[7]. Note that in all these problems, conservation of energy also leads to a constant velocity, but similarly overestimates the numerical coefficient of this velocity[8].

We measured the initial velocity of the liquid column as a function of the depth of immersion $H$. Since the dimensionless law at short time ($t << 6$) just reads $z = t$, introducing dimensional quantities implies a quick variation of the column velocity with $H$. Then, equation 11 just writes:

$$Z(T) = \sqrt{gH}\ T \tag{12}$$

We did experiments with hexane, and found that indeed the height $Z$ of the liquid column increases linearly with time at short time (practically for $t < 1.5$, which corresponds to 10 data points). Thus, we could report its velocity $V$ as a function of the square root of the depth height (Fig. 3), varying $H$ from 2 $cm$ to 35 $cm$, and indeed found a linear relation with a slope $\sqrt{g}$, as predicted by eq. 12. Conservation of energy in eq. 3 for a system starting from $z = 0$ also predicts a regime of constant velocity, but with a higher slope ($\sqrt{2g}$ instead of $\sqrt{g}$). Thus this regime of constant velocity also allows us to stress the existence of an energy loss in this system. Note also that the observed curve does not intercept the origin, which will be shown to be due to entrance effect, caracterized by a length of order $R$. Thus, our model only holds in the limit $H >> R$.

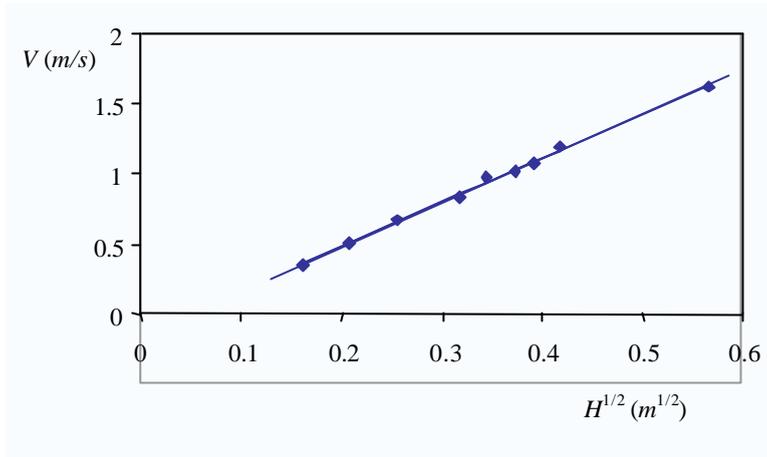

*Figure 3: Rise velocity V of the liquid column at short time (t < 1.5), as a function of the square root of H, the depth of immersion. The full line has a slope $\sqrt{g}$, as expected from eq. 12.*

More generally, eqs. 6a and 6b show that a solution of constant velocity can only be found at the rise and if gravity can be neglected ($dz > 0$ and $z << 1$), or in the case where a horizontal pipe is connected with the bottom of a very large tank, which only generates an entrance flow. On the other hand, eq. 6b shows that the velocity is never constant during the descent, and the only analytical regime in this case is a regime of constant acceleration: leaving a liquid column flow downwards from a very large height ($z_o >> 1$) yields: $\ddot{z} = -1$.

*3.2. Oscillations, and their two regimes of damping*

At longer times, gravity cannot be neglected and eqs. 10a and 10b can be integrated numerically. This solution is drawn in full line in Fig. 4, and compared with data obtained with hexane (for $H = 30$ *cm* and $R = 1$ *cm*).

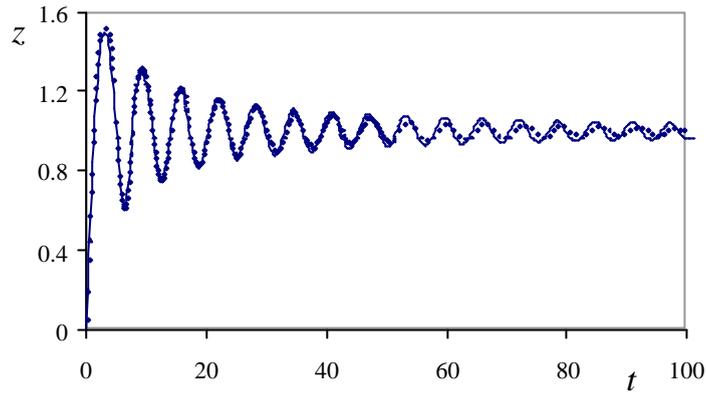

*Figure 4: Height versus time (same normalizations as in Figs. 2 and 3) for a glass tube (radius 10 mm) partially immersed (H = 30 cm) in a large tank filled up with hexane. The dots correspond to experimental data and the full line to a numerical integration of eq. 10.*

The agreement between the theory and the experiment is excellent during the first oscillations: both the positions of the extrema and the periodicity are well predicted by the model. In particular, the first half-oscillation is the parabola derived in eq. 11. After typically ten oscillations, a slight shift appears, and the damping is observed to be quicker than predicted. We interpret this "overdamping" as due to the usual viscous friction along the tube, which must be taken into account as soon as a parabolic Poiseuille-Hagen profile has been established. This is realized after the time necessary for the boundary layer to diffuse on a length $R$, which scales as $\rho R^2/\eta$, with a numerical coefficient of order 0.11, as shown in[9]. This time thus mainly depends on the tube radius, which can be easily checked by doing the same experiment in a thinner tube. Fig. 5 shows the data obtained using a tube twice thinner ($R = 5$ *mm*).

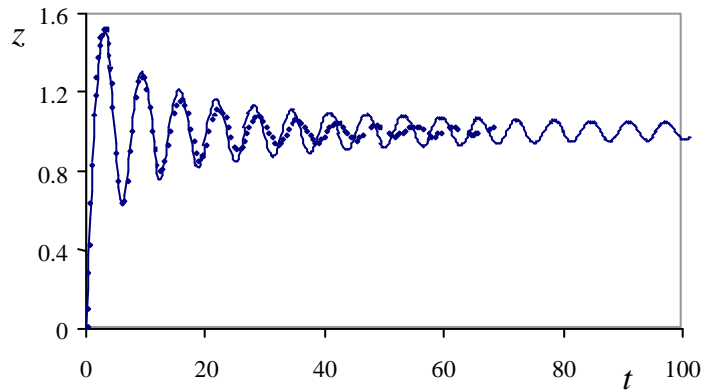

*Figure 5: Same experiment as in Fig. 4, in a thinner tube (R = 5 mm). The dots are the experimental data, while the line corresponds to a numerical integration of eq. 10.*

While the first oscillations remain quite well described by eq. 10, it is indeed observed that the overdamping takes place much earlier: deviations towards eq. 10 are observed around $t = 15$, instead of $t = 60$ (in agreement with the scaling for the time of diffusion of the viscous boundary layer, in $R^2$).

One of the remarkable features of this system is the persistence of the oscillations (typically, more than 20 oscillations can be observed). This is due to the particular source of dissipation in equation 10. If the damping were just caused by the viscous dissipation along the pipe, this would provide a decreasing exponential law for the maxima. In the case we are mainly interested in (short time behavior), the damping is due to the singular pressure loss at the entrance (or exit) of the pipe. The following argument allows us to understand why it is so low. From eqs. 3 and 5, we can derive an equation for the energy loss:

$$\frac{d}{dt}\left(z\dot{z}^2 + (z-1)^2\right) = -|\dot{z}|\dot{z}^2 \tag{13}$$

We set $z(t) = 1 + \alpha(t)\sin t$, with $\alpha \ll 1$, and suppose a slow variation for $\alpha$. During a period, the mean value of the quantities $|\dot{z}|\dot{z}^2$ and $\left(z\dot{z}^2 + (z-1)^2\right)$ are $4|\alpha^3|/3\pi$ and $\alpha^2$, respectively. Thus, an equation for the oscillation amplitude $\alpha$ is obtained from eq. 13:

$$\frac{d\alpha^2}{dt} = -\frac{4|\alpha^3|}{3\pi} \tag{14}$$

which yields:

$$\alpha(t) = \pm\frac{3\pi}{2t} \tag{15}$$

Even if this linear approximation should mainly concern the oscillations of small amplitude, it helps to understand that the damping is unusually long, due to this hyperbolic behavior. Furthermore, an hyperbolic damping is in good agreement with our data even for oscillations of non-negligible amplitude, as shown in Fig. 6 where the maxima and minima corresponding to Fig. 4 are displayed versus time, in a log-log plot.

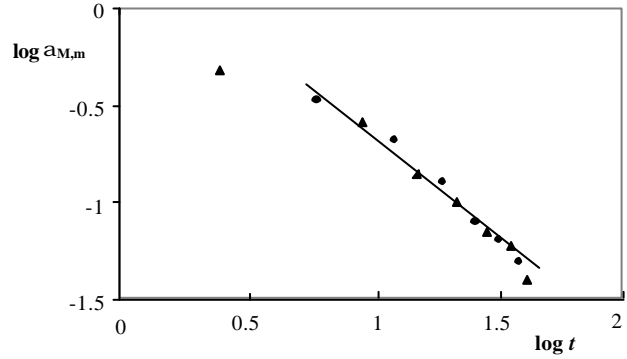

*Figure 6: Successive maxima $a_M$ (triangles) and minima $a_m$ (circles) of the oscillation amplitude as a function of time. The data are taken from Fig. 4 and the full line has a slope –1.*

It is observed that (apart from the first maximum), the damping is close to be hyperbolic in time (the full line indicates the slope –1), before accelerating (the two last points), because of the additional dissipation due to the liquid viscosity. The latter can of course be evaluated by incorporating in the model a viscous Poiseuille friction along the pipe. If the liquid front progresses by a length d$z$, the corresponding energy loss writes (in the same dimensionless variables as above):

$$de = -\Omega z \dot{z} dz \quad (16)$$

where the number $\Omega$ compares viscosity with inertia:

$$\Omega = \frac{16 \eta H^{1/2}}{\rho R^2 g^{1/2}} \quad (17)$$

A difference with the energy loss due to pressure entrance is that d$e$ (energy variation associated with a motion d$z$ of the column) has the same expression whatever the direction of the motion, since $\dot{z}$ and d$z$ have always the same sign. Taking into account this viscous friction modifies eq. 6, which becomes:

$$z\ddot{z} + \dot{z}^2 = 1 - z - \Omega z\dot{z} \quad \text{for } dz > 0 \tag{18a}$$

$$z\ddot{z} = 1 - z - \Omega z\dot{z} \quad \text{for } dz < 0 \tag{18b}$$

Unlike eq. 6, eq. 18 cannot be integrated analytically, but only numerically, except in the case of very large $\Omega$. Then, inertia can be neglected, and the equation for the column motion simply writes:

$$\Omega z\dot{z} = 1 - z \tag{19}$$

which is often referred to, in the context of dynamic capillary rise, as the Washburn equation[10]. At short time (but large enough so that inertia can be neglected), $z$ is small ($z \ll 1$), and integration of eq. 19 shows that the rise follows a diffusion-type law: $z(t) = \sqrt{t/\Omega}$. Then, when approaching equilibrium ($z \to 1$), we find an exponential relaxation: $z(t) = 1 - \exp(-t/\Omega)$.

An interesting feature of eq. 18 is that it allows us to predict if the system will exhibit oscillations, or not. We saw that at large viscosities ($\Omega \gg 1$), the system just relaxes towards equilibrium, without any overshoot of the equilibrium height. Thus, a critical number $\Omega_c$ does exist, below which oscillations develop. Close to $\Omega_c$, we can linearize eq. 18a and 18b, which both give:

$$\ddot{\zeta} + \Omega\dot{\zeta} + \zeta = 0 \tag{20}$$

where we have set: $z = 1 + \zeta$, with $\zeta \ll 1$. This equation only leads to oscillations if $\Omega < \Omega_c = 2$, i.e. for small enough viscosities. Written dimensionally on the depth of immersion, this criterion reads:

$$H < H_c = \frac{\rho^2 g R^4}{64\eta^2} \tag{21}$$

This criterion is largely fulfilled in the series of experiments presented above: with water and centimetric tubes, $H_c$ is of order 1 *km* ! But this height rapidly decreases when making the tube thinner: for a tube of radius 3 *mm* and a liquid 10 times more viscous than water, $H_c$ becomes of order 10 *cm*.

*3.3. Very short time behavior*

*3.3.1 Starting of the liquid column*

Let us come back to the beginning of the rise, starting from $z = 0$. We showed that it obeys a very simple law, since the height of the column increases linearly with time (eq. 12 and Fig. 3). An interesting question is the way the system finds its constant velocity *V*. At $t = 0$, the system is at rest and there is a regime of transition during which the velocity quickly increases from 0 to *V*. Then, the column weight is negligible, and eq. 6a can be written:

$$z\ddot{z} + \dot{z}^2 = 1 \qquad (22)$$

This equation has no solution which verifies both $z = 0$ and $\dot{z} = 0$ for $t = 0$, because of the singularity at z = 0 (then, a zero velocity implies an infinite acceleration for eq. 22 to be obeyed). If we impose a zero initial velocity, we must introduce an initial position $z = a$. Then, eq. 22 can be integrated twice, which yields as a solution close to the origin:

$$z(t) = \sqrt{a^2 + t^2} \qquad (23)$$

This solution has the "good" property to match the solution of constant velocity $z = t$, for *t* larger than *a*. It can also be expanded at very small time ($t << a$), which provides a regime of pure acceleration: $z(t) = a + t^2/2a$. Thus, as emphasized above, the smaller the initial position *a*, the larger the acceleration $1/a$.

By taking pictures at a high rate (typically 1000 frames per second), we could record the very beginning of the rise. Such data are reported in Fig. 7. It is observed that the behavior at a very short time ($t < 0.15$) can indeed be fitted by a parabola of equation:

$z(t) = a_o + t^2/2a$, provided that two different coefficients $a_o$ and $a$ were taken in the equation of the parabola. In the example of Fig. 7, the fit gives: $a_o = 0.064$ and $a = 0.126$. Note that at larger time, the parabolic regime meets the linear one discussed in paragraph 3.1 (eq. 12).

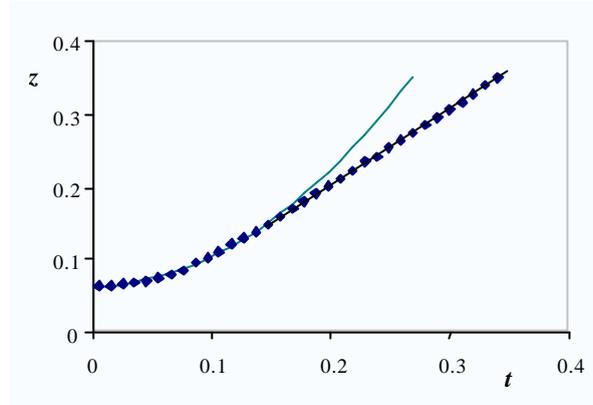

*Figure 7: Height z versus time t, in the very first steps of the rise (R = 20 mm, H = 30 cm and h = 1.9 mm). The data are successively fitted by a parabola of equation $z(t) = a_o + t^2/2a$ (from which the coefficients $a_o$ and a can be deduced) and by a straight line of equation $z(t) = t$.*

The fit can be interpreted in the following way: the coefficient $a_o$ indeed represents the height $z(0) = h/H$ of liquid initially present inside the tube, while $a$ is related to the quantity of liquid initially accelerated. Both are found to be different (with $a > a_o$), which can be related to an effect of added mass: liquid below the tube is also put in motion as the column starts rising. Noting L the equivalent column below the tube, the length initially accelerated is $L + h$. Thus, we can deduce L from the data, since we have $a = (L+h)/H$, or in real units $gH/(L+h)$, where both $h$ and $H$ are known. Because the velocity field in the bath quickly vanishes as a function of the distance to the entrance, we expect L to be of order $R$, the radius of the tube. This is indeed the case, as shown in Fig. 8, where L is plotted versus $R$, and found to vary linearly with it (the numerical coefficient is found to be of order 1).

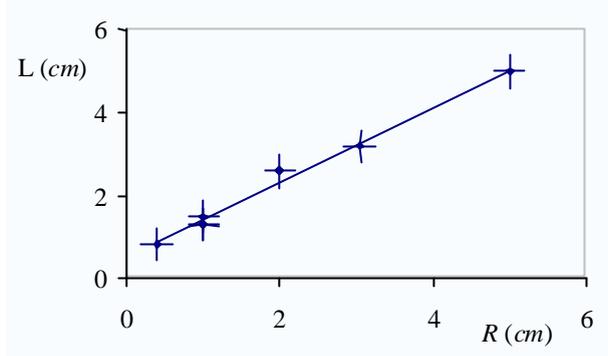

*Figure 8: Entrance length L versus R, the radius of the tube (H = 20 cm and h » 1 cm).*

This result is in close agreement with a prediction by Szekeley *et al.* (in the context of capillarity)[11]. They show by integrating the velocity profile from the entrance of the tube to infinity, that entrance effect can be treated by replacing $Z\ddot{Z}$ in eqs. 6a or 13 by $(Z+7/6R)\ddot{Z}$, which yields as an entrance length: $L = 7/6R$.

We also considered the influence of *h* on L, and focused on the case of an empty tube ($h \rightarrow 0$). Then, as stressed above, the problem should become singular. Practically, it is not; Fig. 9 shows that L does not depend on *h*, which is consistent with the hypothesis of an added mass below the tube entrance. Even in the limit of a tube initially empty ($h \rightarrow 0$), the mass of accelerated fluid is not zero and the acceleration remains finite.

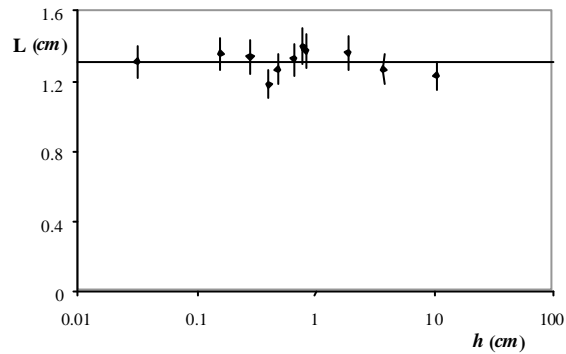

*Figure 9: L versus h plotted for a tube of radius R = 1.2 cm (H = 30 cm).*

Since the flow inside the tube perturbs the reservoir on a length of order *R*, all the conclusions and interpretations presented above (sections 3.1 and 3.2), for which we had $Z \gg R$, remain unchanged. The corrections it induces on the column trajectory only concerns the very first steps of the trajectory, in the accelerating regime illustrated in Fig. 7.

*3.3.2 Jet eruption*

We have up to now focused our discussion on the motion of the whole column, but local deformations of the free surface were also observed at short time. Fig. 10 shows a side view of the tube for *Z* of about 2*R*, where it can be seen that a liquid finger develops at the centre of the tube. This finger first rises and then collapses; we note *A* its maximum height.

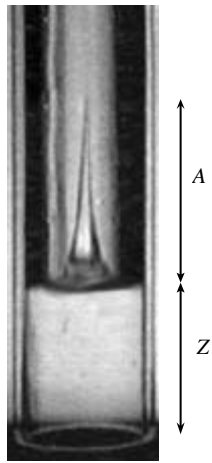

*Figure 10: Early stage of the rise (R = 20 mm, H = 30 cm and h = 0 mm). The front is flat, except at the tube centre where a liquid finger develops.*

The size *A* of this finger depends on the height *h* of liquid initially in the tube, as shown in Fig. 11. Note that *h* can even been made negative, by injecting air bubbles inside the empty pipe, before the rise takes place.

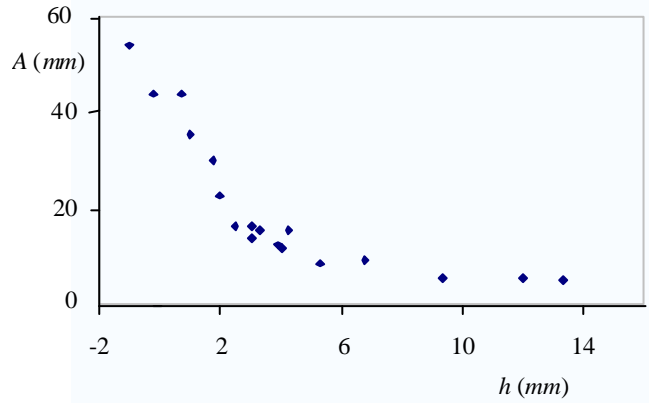

*Figure 11: Maximum amplitude of the liquid finger versus h the initial height of liquid. The experiment was carried out in a tube of radius R = 20 mm and for H = 30 cm.*

For large *h* (> 6 *mm*), the size of the liquid finger does not depend on *h*. In this regime, we still observe some oscillations of the interface due to the abrupt contraction between the reservoir and the tube. Such oscillations were described by Taylor[12], in the case of a tank with an oscillating wall. He showed that free standing waves could set up in the tank, with a shape very close to the one observed in the tube.

For smaller *h*, a strong dependence can be observed: the smaller the height, the longer the finger. We were interested in the dynamics of the finger growth. Fig. 12 reports different series of experiments.

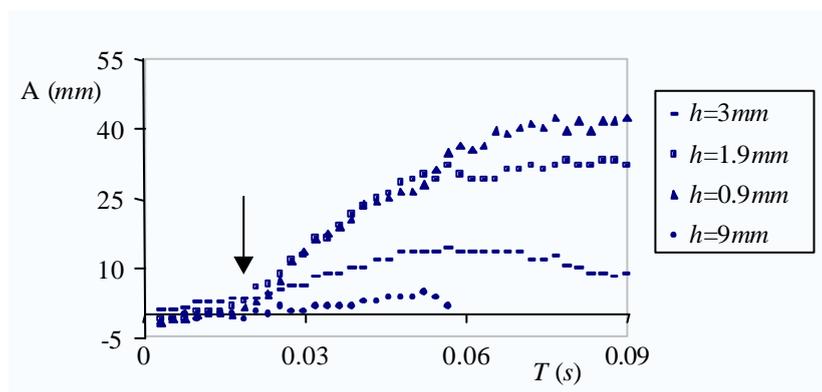

*Figure 12: Amplitude of the finger as a function of time (H = 30 cm and R = 20 mm). The liquid column starts rising at T = 0 and the finger starts developing at the arrow.*

Besides the *h*-dependence of A stressed above, Fig. 12 shows that the finger grows after some delay (typically 0.02 *s*), whatever *h*. This implies that a simple scenario (either a convergence of the flow lines, or a kind of Rayleigh-Taylor instability due to the pulse of acceleration at the beginning of the rise) cannot explain the phenomenon. To go further, we took detailed films of the very beginning of the rise, focusing on the shape of the front interface. A series of snapshots taken at short time from above the tube is displayed in Fig. 13.

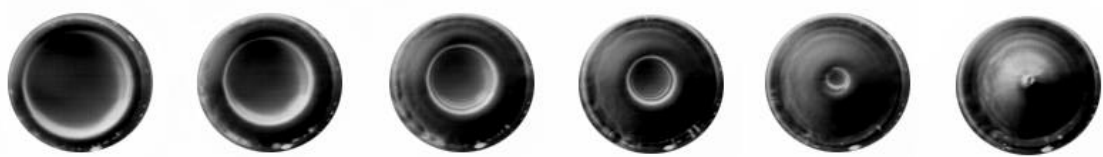

*Figure 13: Set of pictures taken from above each 2.7 ms (R = 20 mm, H = 20 cm and h = 3 mm). These pictures correspond to the first points (before the arrow) in Figure 12. It can be observed that a cavity forms and closes, producing a jet (last picture).*

These pictures show the existence of a circular rim, which sets near the wall of the tube, and closes as time goes on. The collapse of this surface wave produces a jet (last picture of the series), as observed in similar situations[13]. The speed at which the liquid crater closes could be deduced from series similar to Fig. 13. Fig. 14 shows how the diameter $D$ of the liquid crater varies versus time.

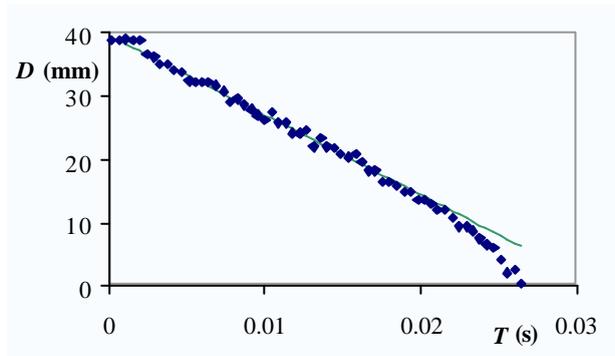

*Figure 14: Diameter of the liquid crater versus the time (R = 20 mm and H = 20 cm).*

The liquid crater closes at a constant velocity, which is 1.3 *m/s* in the above example, of the order of the velocity $\sqrt{gH}$ of the rising column (1.4 *m/s*, in the same experiment). Note that this wave starts propagating during the acceleration phase of the column. We saw that in this phase, the acceleration is of the order of $gH/R$ (for $h = 0$), significantly larger than $g$. Hence, for a wave vector $k$, a typical wave velocity should scale as $\sqrt{gH/kR}$, of the order of $\sqrt{gH}$ for $k \sim 1/R$. The time needed for shutting down the crater in a pipe of radius 20 *mm* is 26 *ms*, of the order of the delay measured in Fig. 12. Then why does this crater establish? The phenomena is probably due to a sticking back of boundary layer and to the associated formation of vortices, revealed in Fig. 15.

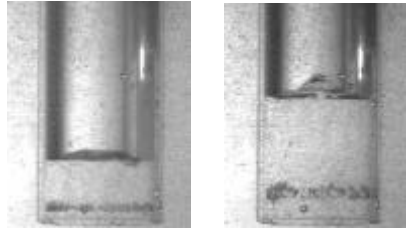

*Figure 15: Visualisation of eddies using small bubbles as a tracer for the flow (R = 20 mm and h = 7 mm).*

The gas bubble ejected from an effervescent drug placed below the tube provide indicators of particle paths. They underscore the existence of a circular vortex ring which stays on the spot while the liquid is filling up the tube. This vortex ring due to the sticking back of the boundary layer deflects the stream line from the edge of the tube, then creating an accumulation of fluid into a circular ring.

If *h* is negative, the initial conditions are different. Then, some air can be trapped in the tube creating a vortex ring of air rather than a liquid one (as seen in the preceding paragraph). This phenomenon could be enhanced by sticking a diaphragm at the tube entrance, as seen in Fig. 16.

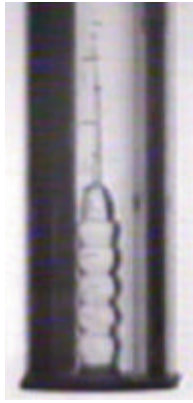

*Figure 16: Liquid column rising in a tube (R = 20 mm) where a diaphragm (of radius 10 mm) partially closes the bottom.*

The column then takes the diameter of the diaphragm, with a modulation of frequency 184 Hz. (Note in addition that the finger previously described is still here, above the main column, as observed in Fig. 16.) The column modulation is probably due to the stationary pressure waves of the air trapped into the tube. The frequency of such a resonant tube (open at one end and closed at the other) is given by the relation:

$$f_n = n\, C/4Y$$

where C is the sound speed, Y the total length of the tube (1.6 *m* in the experiment), and n the mode of oscillation. For n = 1, we find $f_1$ = 187 Hz, in very good agreement with the measured frequency. This agreement remains excellent if the tube length is changed.

## *4. Conclusion*

We have studied the gravitational oscillations of a liquid column initially empty (or nearly empty) and partially immersed inside a large reservoir. We have stressed that this problem has different analytical solutions, depending on the liquid viscosity. For very viscous liquids, the rise should obey the classical laws of impregnation (height proportional to the square root of time, followed by an exponential relaxation towards equilibrium). But the interesting case is the low viscosity limit, for which different features were observed and analyzed, focusing in particular on the first steps of the rise (inertial regimes): after an accelerating phase (where the liquid entrained was mainly the one below the tube), the velocity of rise was found to be a constant fixed by the depth of immersion. Then, the rise was observed to slow down (because of the column weight); we have shown that the trajectory is parabolic, reaching as a first maximum 1.5 times the depth of immersion. This value confirms that indeed energy is dissipated in this inertial phase, because of the sudden contraction endured by the moving liquid which passes from a large reservoir to a finite pipe. After this first maximum, many rebounds were observed, which was understood by evaluating the long range damping associated with this energy loss: the envelop of the height/time dependence was found to be hyperbolic (instead of exponential, as it is the case for usual viscous damping). At long time, viscosity must of course also be considered, which provides a quicker damping of the oscillations. We finally described qualitatively an instability of the liquid/air interface during the first steps of the rise: then, a liquid jet is emitted while the column develops. A complete study of this jet remains to be done.

*Acknowledgement*

We thank Marc Rabaud and Elie Raphael for very valuable comments.